\documentclass[a4paper,11pt]{article}
\usepackage[english]{babel}
\usepackage{amssymb,amsmath}
\usepackage[hang, nooneline]{subfigure}
\usepackage{epsfig,multicol,ifthen,amsfonts,setspace,graphicx,authblk,fancyhdr,everypage}
\usepackage{graphicx}
 
\begin{document}

\title{\bf{Dynamical Symmetry of the Zwanziger problem in Non-commutative Quantum Mechanics}} 

\author[1]{Juhi Rajhans\thanks{libebook@iiserkol.ac.in}}

\affil[1]{Department of Physical Sciences, Indian Institute of Science Education and Research- Kolkata, India}


\pagestyle{fancy}

\maketitle

\thispagestyle{plain}

\section{\bf{Abstract}}
The non-relativistic hydrogen atom and the Zwanziger problem have the same dynamical symmetry for bound and scattering states.We show that this is also true for a Hilbert space which is non-commutative in co-ordinates. The group structure is described using the redefined velocity operator and Laplace Runge-Lenz operator in terms of left and right handed representations of the non-commutative Hilbert space $ R_{\lambda}^{3}$.The bound state algebra is SO(4) and the scattering state algebra is SO(3,1).

\section{Introduction}
Non-commutative quantum field theory is one of the proponent theories for a quantum theory of gravity. The main thought behind this is that at very small distance scales the position co-ordinates do not commute with each other. This leads to breaking of translational invariance and accounts for curvature an could be studied as a theory of quantum gravity. The novel features of NC-QFT are UV-IR mixing which implies that physics at high energies can affect physics at low energies. \\ The fundamental principle of quantum mechanics is the Heisenberg's uncertainty principle. This results in the fuzziness of the phase space. In non-commutative quantum mechanics, the uncertainty principle can be extended to all co-ordinates. Thus even the notion of a single point loses significance. Each such point can be called a fuzzy sphere. Different fuzzy spheres are related in such a way that at large distances the usual flat geometry of $ R^{3}_{0} $ is recovered.

The objective of the article is to introduce the reader to non-commutative quantum mechanics, a simplified version of the much complicated non-commutative quantum field theory .It is hoped that the reader uses this article as a graduate level beginner material for a much more rigorous theory of quantum gravity. The article is the study of the classic Hydrogen atom and a very closely related Zwanziger problem in a modified Hilbert space.The article begins with a preview of the dynamical symmetry of hydrogen atom and the Zwanziger problem, and introduces the reader to non-commutative Hilbert space in 3 dimensions[4] and the preservation of the dynamical symmetry of the coulomb problem. The preservation of the group structure of the Zwanziger problem in the non-commutative Hilbert space is proven. The treatment is non-relativistic and quantum mechanical.  This problem might have a possible application in low-dimensional string theory.
 
 \section{\textbf{The Hydrogen Atom}}

It was shown by Bander and Itzykson[2,3] that the  internal structure of the hydrogen atom bound state can be described by the SO(4) algebra and the the scattering sates can be described by means of SO(3,1) algebra. 
We define 
\begin{center}

 $ \overrightarrow{P} = \frac{\overrightarrow{L} + \overrightarrow{A}}{2} $

 $ \overrightarrow{Q} = \frac{\overrightarrow{L} - \overrightarrow{A}}{2} $

\end{center}

\begin{flushleft}where P and Q are the generators of SO(4) algebra , L is the orbital angular momentum operator and A is the Laplace- Runge Lenz vector.\end{flushleft}

\begin{flushleft}\textbf{SO(4) Algebra}\end{flushleft} \begin{center} $ [P_{i},P_{j}] = i\epsilon_{ijk}P_{k} $\\ $ [Q_{i},Q_{j}] = i\epsilon_{ijk}Q_{k} $ \hspace{2 mm} $[ P_{i},Q_{i}]= [P_{i}, Q_{j}] = 0.$ \end{center} The value of the Casimir operator for each of the SU(2) representations $ P_{i} $ and $ Q_{i} $ - \begin{center} 
$ L.A = 0 = A.L $\\ \ $ P^{2} = Q^{2} = \frac{1}{4}(L^{2} + A^{2}) = \frac{1}{4}\frac{1 + me^{4}}{2E\hbar^{2}} $\end{center} The value of the Casimir gives the value of the energy of the bound states for the hydrogen atom.\begin{center} $ -(n^{2} + 2n) = \frac{1 + me^{4}}{2E\hbar^{2}} $ \hspace{1 mm} $ E = \frac{-me^{4}}{2\hbar^{2}(n + 1)^{2}} $\end{center} 

Similarly, the scattering states of the hydrogen atom can be described by the SO(3,1) algebra.

\subsection{\textbf{ $ H^{3}_{\lambda}$}}

\begin{flushleft}

Following the work of Galikova and Presnajder[4] we study the structure of the non-commutative quantum mechanical Hilbert space $ H^{3}_{\lambda}. $ The non-commutative coordinates in $ R_{0}^{3} $ can be realised in terms of two pairs of creation and annihilation operators[4] $ a_{\alpha}, a^{\dagger}_{\alpha} $ satisfying \begin{center}$ [a_{\alpha},a^{\dagger}_{\beta}]= \delta_{\alpha\beta} $,\hspace{1mm}$ [a_{\alpha},a_{\beta}] = [a^{\dagger}_{\alpha},a^{\dagger_{\beta}}]= 0 .$ \end{center} They act in the auxiliary Fock space spanned by normalised vectors  \begin{center} $ \vert n_{1},n_{2}\rangle = \frac{(a_{1}^{\dagger})^{n_{1}}(a^{\dagger}_{2})^{n_{2}}}{\sqrt(n_{1}!n_{2}!)}\vert 0\rangle $.\end{center} 

 $ \vert0\rangle \equiv \vert0,0\rangle $ denotes the generalised vacuum state: $ a_{1}\vert 0\rangle = \vert a_{2} = 0. $ The auxiliary Fock space is \begin{center} $ F_{n} = \{\vert n_{1},n_{2}\rangle \vert n_{1} + n_{2} = n\} $ \end{center} The non-commutative co-ordinates conform to the rotational symmetry of the Hilbert space and follow SU(2) algebra.\begin{center} $ x_{j} = \lambda a^{\dagger}\sigma_{j}a \equiv \lambda\sigma^{j}_{\alpha \beta} a^{\dagger}_{\alpha} a_{\beta},\hspace{1 mm} j = 1,2,3. $ \end{center} where $ \lambda $ is a universal length parameter and $ \sigma_{j} $ are Pauli matrices.\\ \ 
 The NC analogue of the Euclidean distance - \begin{center}

$ r = \lambda(N+1), \hspace{1 mm} N = a_{\alpha}^{\dagger}a_{\alpha} $. \end{center} The co-ordinates $ x_{j} $ and r satisfy rotationally invariant commutation relations-\begin{center}
$ [x_{i},x_{j}] = 2i\lambda\epsilon_{ijk}x_{k},\hspace{1mm} [x_{i},r]= 0, \hspace{1mm} r^{2}- x_{j}^{2} = \lambda^{2}. $\end{center}

\end{flushleft}

\subsection{\textbf{NC operators}}

In order to study the group structure and symmetry relations, the operators need to be redefined in a non-commutative quantum-mechanical Hilbert space. Thus, following the work of Galikova and Presnajder, we enumerate the redefined operators in $ H^{3}_{\lambda} .$

\begin{flushleft}

 \textbf{Operator Wave Function}: \hspace{2 mm}   $  \Psi = \Sigma C_{m_{1}m_{2}n_{1}n_{2}} (a_{1}^{\dagger})^{m_{1}}(a_{2}^{\dagger})^{m_{2}}(a_{1})^{n_{1}}(a_{2})^{n_{2}},$ 

where the summation is over non-negative finite integers, $ m_{1}+m_{2}= n_{1} +n_{2} $. The Hilbert space possesses the following finite-weighted Hilbert Schmidt norm,\\
$ \vert\vert \Psi \vert\vert^{2} = 4\pi \lambda^{3}Tr[(N+1)\Psi^{\dagger}\Psi] .$ \\ \

 \textbf{Velocity operator}: \hspace{2 mm} $ V_{i} = -i[\hat x_{i},H] $ where $\hat x_{i}\psi = \frac{1}{2}(x_{i}\psi + \psi x_{i}) $

The evolution of the velocity operator is related to that of the co-ordinate operator and is given by the Hamiltonian.\\ \

 \textbf{Angular Momentum}

The generators of rotations is define as follows-\\ $ L_{j}\Psi = \frac{1}{2}[a^{\dagger}\sigma_{j}a,\Psi], j=1,2,3.$\\  The following are the commutation relations and the relations obeyed by the angular momentum operators.\\ $ [L_{i},L_{j}]\Psi \equiv (L_{i}L_{j} - L_{j}L_{i})\Psi = i\epsilon_{ijk}L_{k}\Psi. $\\

With respect to rotations, the doublet of annihilation(creation) operators transforms as a spinor and the triplet of non-commutative co-ordinates transforms as  a vector\begin{center} $ L_{j}\hat a_{\alpha} = -\frac{i}{2}\sigma^{j}_{\alpha\beta}\hat a_{\beta}$,\hspace{1 mm} $ L_{j}\hat a^{\dagger}_{\alpha} = \frac{i}{2}\sigma^{j}_{\beta\alpha}\hat a^{\dagger}_{\beta}  $ \hspace{1 mm} $ L_{i}\hat x_{j} = i\epsilon_{ijk}\hat x_{k} $\end{center}

 \textbf{Laplace operator}: \hspace{2 mm}\begin{center} $ \Delta = -\frac{1}{\lambda \hat r}[\hat a^{\dagger}_{\alpha},[\hat a_{\alpha},\hat \psi]] $\\ where $ \hat r = \lambda (\hat N + 1) $ \end{center}\begin{flushleft} The factor of r in the denominator plays the role in dimensional analysis and the differential operator is described by the double commutator. \end{flushleft}

\textbf{The Laplace Runge-Lenz vector}: \hspace{2 mm} \begin{center} $ \hat A_{k} = \frac{1}{2}\epsilon_{ijk}(\hat L_{i}\hat V_{j} + \hat V_{j}\hat L_{i}) + \frac{q\hat x_{k}}{r} $\\ where $ \hat x_{k}\psi = \frac{1}{2}(x_{k}\psi + \psi x_{k}) $\end{center}\begin{flushleft} The Laplace Runge vector defined here is the symmetric analogue of the Runge Lenz vector defined quantum-mechanically.\end{flushleft}

 \textbf{Auxiliary operators}:\hspace{2 mm}\begin{center} $ \hat a_{\alpha}\psi = a_{\alpha} \psi $,\hspace{2 mm} $ \hat b_{\alpha}\psi = \psi a_{\alpha} $ \hspace{2 mm} $ \hat a^{\dagger} _{\alpha} \psi = a^{\dagger}_{\alpha} \psi $\hspace{2 mm} $ \hat b^{\dagger}_{\alpha}\psi = \psi a^{\dagger}_{\alpha}. $\\ with $ [\hat a_{\alpha},\hat a^{\dagger}_{\beta}] = - [\hat b_{\alpha},\hat b^{\dagger}_{\beta}] = \delta_{\alpha\beta}$ \end{center}

\textbf{The potential term in $ H_{\lambda}$}

\begin{flushleft}The NC analogue of the central potential is the multiplication of the wave function with the potential.\end{flushleft}

\begin{center} 
$  (V\Psi)(r)= V(r)\Psi = \Psi V(r)$\end{center}

 \textbf{NC Laplace equation}: \hspace{2 mm}
\begin{center} $  \Delta_{\lambda}V(r) = 0$ \end{center}

\end{flushleft}

\subsection{\textbf{Dynamical symmetry}}

\subsubsection{\textbf{The Lenz vector conservation}}

The NC Lenz vector commutes with the Hamiltonian $[A_{i},H] = 0.$ 

\begin{center}
$ [\hat A_{i},\hat A_{j}] = i\frac{\omega}{\lambda}(1 + \frac{\omega\lambda}{4})\epsilon_{ijk}\hat L_{k}  $ \\ $[\hat A_{i},\hat A_{j}] = i(-2E + \lambda^{2}E^{2})\epsilon_{ijk}\hat L_{k} $
\end{center}

\begin{flushleft}
There are three independent cases[5]-
\end{flushleft} \begin{itemize}

\item SO(4) symmetry : $ -2E + \lambda^{2}E^{2}> 0 \Longleftrightarrow E<0 $ or $ E> \frac{2}{\lambda^{2}} $,
\item SO(3,1) symmetry : $ -2E + \lambda^{2}E^{2}<0 \Longleftrightarrow 0<E< \frac{2}{\lambda^{2}}$,
\item E(3) Euclidean group: $ -2E + \lambda^{2}E^{2}= 0 \Longleftrightarrow E=0 $ or $ E = \frac{2}{\lambda^{2}} $

\end{itemize}

\begin{flushleft}

The Casimir operators in the above mentioned cases are-$ \hat C'_{1} = \hat L_{j}\hat A_{j} $\hspace{2 mm} $ \hat C'_{2} = \hat A_{i}\hat A_{i} + (-2E + \lambda^{2}E^{2})(\hat L_{i}\hat L_{i} + 1). $Both Casimir operators take constant values $ \hat C'_{1} = 0  $ and $ C'_{2} = q^{2} $ in $ H^{E}_{\lambda} $ since we are dealing with irreducible representations.

\end{flushleft}

\subsubsection{\textbf{Symmetry of The coulomb problem in NC spaces}}
\begin{flushleft}
\textbf{Bound states-SO(4) symmetry}\\

 Rescaling the LRL vector \begin{center} $ \hat K_{j}= \frac{\hat A_{j}}{\sqrt{-2E + \lambda^{2}E^{2}}} $ \end{center}
 
 This generates the representation of the SO(4) algebra.\begin{center}$ [\hat L_{i},\hat L_{j}]= i\epsilon_{ijk}\hat L_{k} $, \hspace{2 mm} $  [\hat L_{i},\hat K_{j}]= i\epsilon_{ijk}\hat K_{k}   $,\\ $ [\hat K_{i},\hat K_{j}]= i\epsilon_{ijk}\hat L_{k} $\end{center}

 The normalised Casimir operators,\begin{center} $ \hat C_{1} = \hat L_{j}\hat K_{j} $ and $ \hat C_{2} = \hat K_{i}\hat K_{i} + \hat L_{i}\hat L_{i} + 1  $ \end{center}

 The second operator equals $ (2j+ 1)^{2} $ for some half integer j and can be written down in terms of energy as\begin{center} $ (2j + 1)^{2} = \frac{q^{2}}{\lambda^{2}E^{2} - 2E} \equiv n^{2} .   $ \end{center}

 The energy eigenvalues $ E = \frac{1}{\lambda^{2}} \mp (\frac{1}{\lambda^{2}}\sqrt{1 + \kappa^{2}_{n}}) $, $ \kappa_{n} = \frac{q\lambda}{n} $[5]

The SO(4)representations which allow for attractive Coulomb potential and repulsive potential at ultra high energies are unitarily equivalent since the Casimir operators take the same values for both cases. At $ \lambda\rightarrow 0 $ the extraordinary bound states at ultra-high energies disappear from the Hilbert space. 

\end{flushleft}

\begin{flushleft} 

\textbf{Scattering states - SO(3,1) symmetry}
\begin{center}
$ \hat K_{j} = \frac{\hat A_{j}}{\sqrt{2E - \lambda^{2}E^{2}}} $ \end{center}

 These yields the SO(3,1) representation group.\begin{center}$ [\hat L_{i},\hat L_{j}]= i\epsilon_{ijk}\hat L_{k} $, \hspace{2 mm} $  [\hat L_{i},\hat K_{j}]= i\epsilon_{ijk}\hat K_{k}   $,\\ $ [\hat K_{i},\hat K_{j}]= - i\epsilon_{ijk}\hat L_{k} $ \end{center}

\end{flushleft}

\section{\textbf{ The symmetries and group structure of the Zwanziger problem}}

It was observed that very interesting similarities of symmetries existed between a hydrogen atom and interactions between particles having both electric and magnetic charges.It is found that the Dirac quantisation condition leads to two elementary units of charge[1].
The charge-monopole problem exhibits the same higher symmetry as the hydrogen atom. The Hamiltonian of the Zwanziger problem is given by \begin{center}$ H = \frac{\pi^{2}}{2m} - \frac{\gamma}{r} + \frac{\mu^{2}}{2mr^{2}}. $\end{center} where u and v are velocities of the charged particle for the bound states and scattering states respectively.

\begin{center}

\begin{tabular}{|c|c|}

\hline 
Scattering state- SO(3,1)Algebra & Bound State-SO(4)) \\ 
\hline 
$  K = \frac{A}{v}, \gamma' = \gamma/v $ & $ N = \frac{A}{u}, \gamma' = \gamma/u $ \\ 
\hline 
$ [J_{i},J_{j}]= i\epsilon_{ijk}J_{k}$ & $ [J_{i},J_{j}]= i\epsilon_{ijk}J_{k}$\\ 
\hline 
$ [J_{i},K_{j}]= i\epsilon_{ijk}K_{k}$ & $ [J_{i},N_{j}]= i\epsilon_{ijk}N_{k}$ \\ 
\hline 
$ [K_{i},K_{j}] =- i\epsilon_{ijk}J_{k}$ & $ [N_{i},N_{j}] = i\epsilon_{ijk}J_{k} $ \\ 
\hline 
 $ J^{2} - K^{2} = ( \mu^{2} - \alpha'^{2} - 1) $&  $ J^{2} + N^{2} = (\gamma'^{2} + \mu^{2} - 1) $ \\ 
\hline 

\end{tabular} 

\end{center}

Thus the Hydrogen atom symmetry is conserved even for interaction of particles having both electric and magnetic charges[1].

\section{\textbf{Zwanziger Problem group structure in NCQM}}

In NCQM, the Zwanziger problem group structure is expected to be the same as that in the usual quantum mechanical Hilbert space. The representation space in the NC Hilbert space has left and right creation and annihilation operators. It is known that the non-commutative quantum Hilbert space  is described by left and right  representations has a commutative Hilbert subspace.
\begin{center}
 $ \hat a_{\alpha}\psi = a_{\alpha} \psi $,\hspace{2 mm} $ \hat b_{\alpha}\psi = \psi a_{\alpha} $ \hspace{2 mm} $ \hat a^{\dagger} _{\alpha} \psi = a^{\dagger}_{\alpha} \psi $\hspace{2 mm} $ \hat b^{\dagger}_{\alpha}\psi = \psi a^{\dagger}_{\alpha}. $\\ with $ [\hat a_{\alpha},\hat a^{\dagger}_{\beta}] = - [\hat b_{\alpha},\hat b^{\dagger}_{\beta}] = \delta_{\alpha\beta}$ \end{center}

One defines \begin{center} $ c_{\alpha} = \frac{a_{\alpha} + b_{\alpha}}{2} $, \hspace{2 mm} $ x_{ic} = \sigma^{i}_{\alpha\beta}c^{\dagger}_{\alpha}c_{\beta}. $\end{center} One can check that the commutator of the position operators $ [\hat x_{ic},\hat x_{jc} = 0]. $\\ \  Hence $ x_{c} $ forms a representation for a commutative sub-algebra[6] $ A_{0} $ within the non-commutative algebra $ A_{\theta} $. The velocity operator is $ v_{ic} = -[\hat x_{ic},H] .$ The Hamiltonian is the same as the original Hamiltonian and Laplace Runge Lenz vector with $ x_{c} $ and $ r_{c}$ instead of x and r. The distance coordinate $ r_{c} = \Sigma x_{ic}x_{ic} $ Thus for the non-commutative Hilbert space,the magnetic field due to a monopole at the origin is \begin{equation}
 B(r_{c}) = (\frac{g}{4\pi})\frac{\hat r_{c}}{r_{c}^{2}} 
\end{equation}
and the vector potential for B-
\begin{equation}
\hat A(r_{c}) = \frac{g}{4\pi}\frac{r_{c}\times \hat n r_{c}.\hat n}{r_{c}[r_{c}^{2} - (r_{c}.\hat n)^{2}]}, \hspace{3mm} \hat A_{kc} = \frac{1}{2}\epsilon_{ijk}(\hat L_{ic}\hat V_{jc} + \hat V_{jc}\hat L_{ic}) + \frac{q\hat x_{kc}}{r_{c}} 
\end{equation} 

where$ \hat A(r_{c})$ has singularities along the line and the angular momentum operator $ L_{jc}\Psi = \frac{1}{2}[c^{\dagger}\sigma_{j}c,\Psi] $. 

Now to study the interaction between two particles with masses $ m_{1} $ and $ m_{2}$ charges $ e_{1}$ and $ e_{2} $ and magnetic charges $ g_{1}$ and $ g_{2} $ , the Hamiltonian H is given by 

\begin{center}

 $ H = \frac{1}{2m_{1}}(p_{1c} - \frac{e_{1}g_{2} - e_{2}g_{1}}{4\pi c}D(r_{1c} - r_{2c}))^{2}$\\ \ \\ $ + \frac{1}{2m_{2}}(p_{2c} - \frac{e_{2}g_{1} - e_{1}g_{2}}{4\pi c}D(r_{2c} - r_{1c}))^{2} + \frac{e_{1}e_{2} + g_{1}g_{2}}{4\pi}\frac{1}{|r_{1c} - r_{2c}|} + V(r_{1c} - r_{2c}) $

\end{center}

This leads to the following equations of motion consisting of a Lorentz force, a Coulomb force and a force produced by the Biot-Savart fields. The interaction preserves PT symmetry but violates each of them separately. The Hamiltonian in the centre of mass and relative co-ordinates frame is-
\begin{equation}
 H = \frac{P^{2}_{c}}{2m} + \frac{1}{2m}[p - \mu D(r_{c})]^{2} + V(r_{c}) - \gamma/r_{c}
\end{equation}

which can be re-expressed to a manifestly gauge invariant and rotationally invariant form-

\begin{equation}
H = \pi^{2}_{c}/2m - \gamma/r_{c} + V(r_{c}) 
\end{equation}

\begin{equation}
[x_{ic},x_{jc}] = 0, \hspace{1mm} [\pi_{ic},x_{jc}] = -i\delta_{ij}
\end{equation}

\begin{equation} 
[\pi_{ic},\pi_{jc}] = i\mu\epsilon_{ijk}x_{kc}/r^{3} 
\end{equation}

The commutation relations generated by the angular momentum and the co-ordinates generate a Lie algebra of the three dimensional similarity group $ S_{3} $ which consists of dilations and the three dimensional euclidean group $ E_{3} $ which generates rotations and translations. The irreducible representations of this group are labelled by the quantity $ \hat x_{c}.J_{c} $ which takes the following values- 

\begin{equation}
\hat x_{c}.J_{c} = -\mu = 0,\pm \frac{1}{2}, \pm 1... 
\end{equation} 
 
which is an algebraic derivation of the Dirac quantisation condition. Rewriting the Hamiltonian as 

\begin{equation}
H = \frac{\pi_{rc}^{2}}{2m} + \frac{1}{2mr^{2}_{c}}(J^{2}_{c} - \mu^{2}) - \frac{\gamma}{r_{c}} + V(r_{c}) 
\end{equation}

Imposing $ V(r_{c}) = \frac{\mu^{2}}{2mr^{2}_{c}} $, we find that the resulting problem has a similar form as the hydrogen atom problem or the Coulomb problem.There is no possibility of bound states unless $ \gamma > 0.$ Thus taking the energy to be less than zero, one can write down the commutation relations as the following-

\begin{equation}
N_{c} = \frac{A_{c}}{u_{c}}, \hspace{2mm} \gamma' = \frac{\gamma}{u_{c}} 
\end{equation}

\begin{equation}
[J_{ic},J_{jc}]= i\epsilon_{ijk}J_{kc}, \hspace{3mm} [J_{ic},N_{jc}]= i\epsilon_{ijk}N_{kc} 
\end{equation}

\begin{equation}
[N_{ic},N_{jc}] = i\epsilon_{ijk}J_{kc}, \hspace{3 mm} J^{2}_{c} + N^{2}_{c} = (\gamma'^{2} + \mu^{2} - 1)  
\end{equation}

The above equations give the defining algebra of $ SU(2)\otimes SU(2)\sim O(4).$ One introduces

\begin{equation}
J_{c\pm } = \frac{1}{2}(J_{c}\pm N_{c}) 
\end{equation}
which yields

\begin{equation}
[J_{ic\pm},J_{jc\pm}] = i\epsilon_{ijk}J_{kc\pm}, \hspace{2mm} [J_{ci+},J_{cj-}] = 0,\hspace{2 mm}J_{c\pm}^{2} = \frac{1}{4}[(\gamma' \pm \mu)^{2} - 1]
\end{equation}

This gives the following relations-

\begin{equation}
j_{>} + j_{<} + 1 = \gamma', \hspace{3 mm} j_{>} - j_{<} = |\mu|
\end{equation}

\begin{center}

\begin{tabular}{|c|c|}

\hline 
Scattering state- SO(3,1)Algebra & Bound State-SO(4)) \\ 
\hline 
$  K_{c} = \frac{A_{c}}{v_{c}}, \gamma' = \gamma/v_{c} $ & $ N _{c}= \frac{A_{c}}{u_{c}}, \gamma' = \gamma/u_{c} $ \\ 
\hline 
$ [J_{ic},J_{jc}]= i\epsilon_{ijk}J_{kc}$ & $ [J_{ic},J_{jc}]= i\epsilon_{ijk}J_{kc}$\\ 
\hline 
$ [J_{ic},K_{jc}]= i\epsilon_{ijk}K_{kc}$ & $ [J_{ic},N_{jc}]= i\epsilon_{ijk}N_{kc}$ \\ 
\hline 
$ [K_{ic},K_{jc}] =- i\epsilon_{ijk}J_{kc}$ & $ [N_{ic},N_{jc}] = i\epsilon_{ijk}J_{kc} $ \\ 
\hline 
 $ J^{2}_{c} - K^{2}_{c} = ( \mu^{2} - \gamma'^{2} - 1) $&  $ J^{2}_{c} + N^{2}_{c} = (\gamma'^{2} + \mu^{2} - 1) $ \\ 
\hline 

\end{tabular} 

\end{center}

\section{Conclusions} 

Thus the group symmetry for the Coulomb problem and the Zwanziger problem bound states and scattering states in non-commutative quantum mechanics is the same.It is known that String theory deals with  strings and branes which have electric and magnetic charges.This result could have applications in low-dimensional string theory which is a non-commutative quantum field theory.

\section{Acknowledgements}

The author thanks all those who helped her in this work. She acknowledges financial support from Kishore Vaigyanik Protsahan Yojana fellowship, DST India. She also acknowledges guidance from Prof. Prasanta Panigrahi(IISER Kolkata) and Prof. Sachindeo Vaidya. This work has been possible solely due to their engaging scientific discussions and guidance.

\section{References}


\end{document}